\documentclass[12pt,cite,epsf,epsfig]{article}
\usepackage{epsfig}

\begin{document}
\author{S. Dev\thanks{dev5703@yahoo.com} $^{,1}$,
Sanjeev Kumar\thanks{sanjeev.verma@thapar.edu} $^{,2}$, Surender
Verma\thanks{ s\_7verma@yahoo.co.in} $^{,1}$ and Shivani
Gupta\thanks{shiroberts\_1980@yahoo.co.in} $^{,1}$}
\title{Phenomenological Implications of a Class of Lepton Mass Matrices}
\date{$^1$\textit{Department of Physics, Himachal Pradesh University, Shimla 171005, INDIA.}\\
\smallskip
$^2$\textit{School of Physics and Material Science, Thapar
University, Patiala 147004, INDIA.}}

\maketitle

\begin{abstract}
Phenomenological implications of a class of lepton mass matrices
with parallel texture structure have been examined and
phenomenologically interesting constraints on charged lepton and
neutrino mass matrix parameters have been obtained.
\end{abstract}

\section{INTRODUCTION}
Mass matrices provide important tools for the investigation of the
underlying symmetries and the resulting dynamics. The first step
in this direction is the reconstruction of the neutrino mass
matrix in the flavor basis. However, the reconstruction results in
a large variety of possible structures of mass matrices depending
strongly on the mass scale, mass hierarchy and the Majorana
phases. In the flavor basis, the mass matrix for Majorana
neutrinos contains nine physical parameters viz. three mass
eigenvalues, three mixing angles and the three CP-violating
phases. The two squared-mass differences ($\Delta m^2_{12}$ and
$\Delta m^2_{13}$) and the two mixing angles ($\theta_{12}$ and
$\theta_{23}$) have been measured in solar, atmospheric and
reactor experiments. The third mixing angle $\theta_{13}$  and the
Dirac-type CP-violating phase $\delta$ are expected to be measured
in the forthcoming neutrino oscillation experiments. The possible
measurement of the effective Majorana mass in neutrinoless double
â decay searches will provide an additional constraint on the
remaining three neutrino parameters viz. the neutrino mass scale
and two Majorana CP-violating phases. While the neutrino mass
scale will be independently determined by the direct beta decay
searches and cosmological observations, the two Majorana phases
will not be uniquely determined even if the absolute neutrino mass
scale is known. Thus, it is not possible to fully reconstruct the
neutrino mass matrix with the observations from the presently
feasible experiments. Under the circumstances, it is natural to
employ other theoretical inputs for the reconstruction of the
neutrino mass matrix. The possible forms of these additional
theoretical inputs are constrained by the existing neutrino data.
Several proposals have been made in the literature to restrict the
form of the neutrino mass matrix by reducing the number of free
parameters which include presence of texture zeros
\cite{1,2,3,4,5,6,7,8}, requirement of zero determinant \cite{9,10}
and the zero trace condition \cite{11} amongst others. Some attempts
aimed at understanding the pattern of quark/lepton masses and
mixings by introducing Abelian or non-Abelian flavor symmetries,
naturally, lead to texture zeros in the mass matrices. However,
the current low energy data are consistent with a limited number
of texture zero schemes \cite{1,2,3,4,5,6,7,8}. In particular, the
latest low energy data disfavor \cite{12} all quark mass matrices
with five or more texture zeros. Four texture zero Ans\"{a}tze is
specially important since it can successfully describe not only
the quark but also the lepton sector including the charged lepton
and neutrino masses. Furthermore, not all of the texture zeros in
the four texture zero Ans\"{a}tze can be obtained through weak
basis transformations and, therefore, have physical implications.
In the present work, we examine a special case of four texture
zero Ans\"{a}tze in which the charged lepton mass matrix and
Majorana neutrino mass matrix have the same texture zero structure
with zero entries at (1,1) and (1,3) places. This Ans\"{a}tze has
been analyzed earlier \cite{13} under the assumption of
factorizable phases in the neutrino mass matrix which, however, is
not always possible for a general complex symmetric matrix without
unnatural fine tuning of the phases. In the context of the Type II
see-saw mechanism for example, the phases of the Dirac neutrino
mass matrix and the right handed Majorana neutrino mass matrix are
assumed to be fine tuned to achieve the cancellation of phases in
most of the analyses \cite{14,15} (and references therein) reported
hitherto. In this sense, the analyses reported so far are
incomplete. The condition of factorizable phases in the neutrino
mass matrix was relaxed in \cite{16}. However, this analysis was
confined to the investigation of neutrino hierarchy and effective
Majorana mass. In the present work, we have attempted an
exhaustive analysis of a class of four-zero texture lepton mass
matrices by examining its full implications for the lepton mass
matrices.

All the information about lepton masses and mixings is encoded in
the  hermitian charged lepton mass matrix $M_l$ and the complex
symmetric neutrino mass matrix $M_{\nu}$. In the present work, we
consider a special case of four texture zero Ans\"{a}tze in which
the charged lepton mass matrix and Majorana neutrino mass matrix
have the same texture zero structure with zero entries at (1,1)
and (1,3) places and are given by
\begin{equation}
M_l=\left(%
\begin{array}{ccc}
  0 & A_l & 0 \\
  A_l & D_l & B_l \\
  0 & B_l & C_l \\
\end{array}%
\right)
\end{equation}
and
\begin{equation}
M_\nu=\left(%
\begin{array}{ccc}
  0 & A_\nu & 0 \\
  A_\nu& D_\nu & B_\nu \\
  0 &B_\nu  & C_\nu \\
\end{array}
\right)
\end{equation}
respectively. The hermiticity of $M_l$ requires its diagonal
elements $D_l$ and $C_l$ to be real whereas the non-diagonal
elements $A_l$, $B_l$ are, in general, complex with $A_l=
|A_l|e^{i\phi_1}$, $B_l= |B_l|e^{i\phi_2}$. In contrast, all the
non-vanishing elements of the complex symmetric matrix $M_{\nu}$
are, in general, complex.

The hermitian mass matrix, $M_l$ for the charged lepton, is
diagonalized by the unitary transformation
\begin{equation}
M_l=V_lM_l^dV_l^\dagger,
\end{equation}
where $V_l^\dagger=V_l^{-1}$. The hermitian matrix $M_l$ can, in
general, be written as
\begin{equation}
M_l=P_lM_l^{r}P_l^\dagger
\end{equation}
where $P_l$ is a unitary diagonal phase matrix,
$diag(e^{i\phi_1},1,e^{i\phi_2})$, and $M_l^{r}$ is real matrix
which can be diagonalized by a real orthogonal matrix $O_l$:
\begin{equation}
M_l^{r}=O_lM_l^{d}O_l^T
\end{equation}
where the superscript T denotes transposition and
$M_l^d$=$diag(m_e,-m_{\mu}, m_{\tau})$. From Eqs.(3-5), the unitary
matrix $V_l$ is given by
\begin{equation}
V_l=P_lO_l.
\end{equation}
Using the invariants, Tr$M_l^{r}$, Tr$M_l^{r^ 2}$ and
Det$M_l^{r}$, the matrix elements $|A_l|$, $|B_l|$ and $C_l$ can
be written in terms of the charged lepton masses and $D_l$ as
\begin{eqnarray}
C_l=m_e-m_{\mu}+m_{\tau}-D_l,\nonumber\\
|A_l|=\left(\frac{m_em_{\mu}m_{\tau}}{C_l}\right)^{\frac{1}{2}},\\
|B_l|=\left[\frac{(m_{\tau}-m_{\mu}-D_l)(m_{\tau}+m_e-D_l)(m_{\mu}-m_e+D_l)}{C_l}\right]^{\frac{1}{2}}.\nonumber
\end{eqnarray}
Here, $D_l$ should be in the range
$(m_e-m_\mu)<D_l<(m_\tau-m_\mu)$ for the elements $|A_l|$ and
$|B_l|$ to be real.

Using Eq. (7), the elements of the diagonalizing matrix, $O_l$
can be written in terms of charged lepton masses $m_e$, $m_{\mu}$,
$m_{\tau}$ and charged lepton mass matrix element $D_l$, which are
given by
\begin{equation}
 \left.\begin{array}{c}

O_{11}=\sqrt{\frac{m_{\mu}m_{\tau}(m_{\tau}-m_{\mu}-D_l)}{(m_e-m_{\mu}+m_{\tau}-D_l)
(m_e+m_{\mu})(m_{\tau}-m_e)}}
\\

O_{12}=\sqrt{\frac{m_{\tau}m_e(m_{\tau}+m_e-D_l)}{(m_e-m_{\mu}+m_{\tau}-D_l)
(m_e+m_{\mu})(m_{\tau}+m_{\mu})}}   \\

O_{13}=\sqrt{\frac{m_em_{\mu}(m_{\mu}-m_e+D_l)}{(m_e-m_{\mu}+m_{\tau}-D_l)
(m_{\mu}+m_{\tau})(m_{\tau}-m_e)}}   \\

O_{21}=\sqrt{\frac{m_e(m_{\tau}-m_{\mu}-D_l)}{(m_e+m_{\mu})(m_{\tau}-m_e)}}   \\

O_{22}=\sqrt{\frac{m_{\mu}(m_{\tau}+m_e-D_l)}{(m_{\mu}+m_{\tau})(m_e+m_{\mu})}} \\

O_{23}=\sqrt{\frac{m_{\tau}(m_{\mu}-m_e+D_l)}{(m_{\tau}-m_e)(m_{\mu}+m_{\tau})}} \\

O_{31}=\sqrt{\frac{m_e(m_{\mu}-m_e+D_l)(m_{\tau}+m_e-D_l)}{(m_e-m_{\mu}+m_{\tau}-D_l)
(m_{\mu}+m_e)(m_{\tau}-m_e)}}   \\

O_{32}=\sqrt{\frac{m_{\mu}(m_{\mu}-m_e+D_l)(m_{\tau}-m_{\mu}-D_l)}{(m_e-m_{\mu}+m_{\tau}-D_l)
(m_{\tau}+m_{\mu})(m_e+m_{\mu})}}   \\

O_{33}=\sqrt{\frac{m_{\tau}(m_{\tau}-m_{\mu}-D_l)(m_{\tau}+m_e-D_l)}{(m_e-m_{\mu}+m_{\tau}-D_l)
(m_{\mu}+m_{\tau})(m_{\tau}-m_e)}}

\end{array}  \right\}
\end{equation}
If $D_l$ is known the diagonalizing matrix $O_l$ and the charged
lepton mass matrix $M_l^r$ are fully determined since the charged
lepton masses are known.

The complex symmetric neutrino mass matrix, $M_{\nu}$ is
diagonalized by an orthogonal matrix, $V_{\nu}$
\begin{equation}
M_\nu=V_\nu M_ \nu^{diag}V_\nu^T.
\end{equation}
The neutrino mixing matrix or Pontecorvo-Maki-Nakagawa-Sakata
matrix, $U_{PMNS}$, is given by
\begin{equation}
U_{PMNS}=V_l^\dagger V_\nu.
\end{equation}
The neutrino mixing matrix $U_{PMNS}$ consists of three non
trivial CP-violating phases: the Dirac phase, $\delta$, and two
Majorana phases, $\alpha$ and $\beta$, and, three neutrino mixing
angles viz. $\theta_{12}$, $\theta_{23}$ and $\theta_{13}$. The
neutrino mixing matrix can be written as product of two matrices
characterizing Dirac and Majorana type CP violation
\begin{equation}
U_{PMNS}=UP
\end{equation}
where $U$ and $P$ are given by
\begin{equation}
U=\left(
\begin{array}{ccc}
c_{12}c_{13} & s_{12}c_{13} & s_{13}e^{-i\delta } \\
-s_{12}c_{23}-c_{12}s_{23}s_{13}e^{i\delta } &
c_{12}c_{23}-s_{12}s_{23}s_{13}e^{i\delta } & s_{23}c_{13} \\
s_{12}s_{23}-c_{12}c_{23}s_{13}e^{i\delta } &
-c_{12}s_{23}-s_{12}c_{23}s_{13}e^{i\delta } & c_{23}c_{13}
\end{array}
\right), P = \left(
\begin{array}{ccc}
1 & 0 & 0 \\
0 & e^{i\alpha } & 0 \\
0 & 0 & e^{i\left( \beta +\delta \right) }
\end{array}
\right).
\end{equation}
Using Eqs. (9) and (10), the neutrino mass matrix $M_\nu$ can be
written as
\begin{equation}
M_\nu=P_lO_lU_{PMNS}M_\nu^{diag}U_{PMNS}^TO_l^TP_l.
\end{equation}
The two zero textures in $M_\nu$ yield two complex equations viz.
\begin{equation}
m_1 a^2+m_2 b^2e^{2i\alpha}+m_3c^2e^{2i(\beta+\delta)}=0
\end{equation}
and
\begin{equation}
m_1ad+m_2bge^{2i\alpha}+m_3che^{2i(\beta+\delta)}=0
\end{equation}
where the complex coefficients $a$, $b$, $c$, $d$, $g$ and $h$ are
given by
\begin{eqnarray}
a=O_{11}U_{e1}+O_{12}U_{m1}+O_{13}U_{t1},\nonumber\\
b=O_{11}U_{e2}+O_{12}U_{m2}+O_{13}U_{t2},\nonumber\\
c=O_{11}U_{e3}+O_{12}U_{m3}+O_{13}U_{t3},\nonumber\\
d=O_{31}U_{e1}+O_{32}U_{m1}+O_{33}U_{t1},\nonumber\\
g=O_{31}U_{e2}+O_{32}U_{m2}+O_{33}U_{t2},\nonumber\\
h=O_{31}U_{e3}+O_{32}U_{m3}+O_{33}U_{t3}.\nonumber\\
\end{eqnarray}
Solving Eqs. (14-15) for the two mass ratios
$\left(\frac{m_1}{m_2}\right)$ and $\left(\frac{m_1}{m_3}\right)$,
we obtain
\begin{equation}
\begin{array}{c}
\frac{m_1}{m_2}e^{-2i\alpha}=\frac{b(cg-bh)}{a(ah-cd)},\\
\frac{m_1}{m_3}e^{-2i\beta}=\frac{c(bh-cg)}{a(ag-bd)}e^{2i\delta}.
\end{array}
\end{equation}

One can enumerate the number of parameters in Eqs. (17). The nine
parameters (three neutrino mixing angles
$(\theta_{12},\theta_{23},\theta_{13})$, three neutrino mass
eigenvalues $(m_1,m_2,m_3)$, two Majorana phases $(\alpha,\beta)$
and one Dirac-type CP violating phase, $\delta$) come from
neutrino sector and four parameters (three charged lepton masses
$(m_e,m_\mu,m_\tau)$ and $D_l$) come from charged lepton sector,
thus, totalling 13 parameters. The three charged lepton masses are
\cite{17}
\begin{equation}
m_e=0.510998910 MeV, \\
m_{\mu}=105.658367 MeV, \\
m_{\tau}=1776.84 MeV. \\
\end{equation}
The experimental constraints on the neutrino parameters  at one
standard deviation are \cite{18}
\begin{equation}
\begin{array}{c}
\Delta m_{12}^{2} =7.67_{-0.21}^{+0.22}\times 10^{-5}eV^{2},
 \\
\Delta m_{13}^{2} = (2.37\pm{0.15})\times 10^{-3}eV^{2},
 \\
\theta_{12} =34.5\pm{1.4},   \\
\theta_{23} =42.3_{-3.3}^{+5.1}.
\end{array}
\end{equation}
Only an upper bound is known on the mixing angle $\theta_{13}$
from the CHOOZ experiment. However, the latest global analysis
\cite{19} gives
\begin{equation}
\sin^2\theta_{13}=0.016_{-0.01}^{+0.01}
\end{equation}
which is non-zero at $90\%$ C.L.. The parameters $\delta$ and
$D_l$ are varied uniformly within their full ranges. Thus, we are
left with three unknown parameters viz. $m_1$, $\alpha$, $\beta$.

From Eqs. (17), two mass ratios $(\frac{m_1}{m_2})$ and
$(\frac{m_1}{m_3})$ can be written as
\begin{equation}
\frac{m_1}{m_2}=\left|\frac{b(cg-bh)}{a(ah-cd)}\right|,
\end{equation}
\begin{equation}
\frac{m_1}{m_3}=\left|\frac{c(bh-cg)}{a(ag-bd)}e^{2i\delta}\right|.
\end{equation}
For the simultaneous existence of two texture zeros at $(1,1)$ and
$(1,3)$ positions in $M_{\nu}$, the two values of $m_1$ given by
\begin{equation}
m_{1}=\left(\frac{m_1}{m_2}\right) \sqrt{\frac{ \Delta
m_{12}^{2}}{1-\left(\frac{m_1}{m_2}\right) ^{2}}}
\end{equation}
and
\begin{equation}
m_{1}=\left(\frac{m_1}{m_3}\right) \sqrt{\frac{\Delta m_{12}^{2}+
\Delta m_{23}^{2}}{ 1-\left(\frac{m_1}{m_3}\right)^{2}}},
\end{equation}
calculated from the mass ratios
$\left(\frac{m_1}{m_2},\frac{m_1}{m_3}\right)$, respectively, must
be identical. This constraint can be used to constrain the unknown
variables $\delta$, $D_l$ and the third mixing angle $\theta_{13}$
which has not been measured experimentally as yet. We have, also,
calculated the effective Majorana mass $M_{ee}$ appearing in the
neutrinoless double beta decay for the allowed parameter space
\begin{equation}
M_{ee}=m_1|U_{e1}|^2+m_2|U_{e2}|^2+m_3|U_{e3}|^2.
\end{equation}
The Majorana phases $\alpha$ and $\beta$ are given by
\begin{equation}
\alpha=-\frac{1}{2}arg\left(\frac{b(cg-bh)}{a(ah-cd)}\right),
\end{equation}
\begin{equation}
\beta=-\frac{1}{2}arg\left(\frac{c(bh-cg)}{a(ag-bd)}\times
e^{2i\delta}\right).
\end{equation}
Using these relations, the Majorana phases $\alpha$ and $\beta$
have been calculated for the allowed values of $\delta$,
$\theta_{13}$ and $D_l$.

\section{Results and Discussion}
The $\delta$, $\theta_{13}$ and $D_l$ parameter space allowed by
the current neutrino oscillation data have been depicted in Fig.
1(a) and 1(b) as two dimensional scatter plots. We see that there
are two distinct solutions in the ($\delta$, $\theta_{13}$, $D_l$)
parameter space. The more probable solution corresponds to lower
values of $D_l$ and will be called "low $D_l$" solution,
henceforth. The less probable solution corresponds to higher
values of $D_l$ and will be referred to as the "high $D_l$"
solution in the following discussion. In Fig. 1(c) and 1(d)
$M_{ee}-D_l$ and $\alpha-\beta$ correlation plots have been
depicted. It is evident from Fig. 1 that these two solutions of
$D_l$ have distinguishing implications, especially, for
$\theta_{13}$. One of the characterizing feature of low $D_l$
solution is the existence of lower bound on $\theta_{13}$ i.e.
$\theta_{13}>2.8^o$. The 95\% C.L. ranges of $\theta_{13}$ and
$D_l$ for these two solutions have been given in Table 1. Majorana
mass term $M_{ee}$ is sharply constrained having values of the
order of $10^{-3}$ for low $D_l$ solution, However, for high $D_l$
a wide range of $M_{ee}$ is allowed. The 95\% C.L. ranges of
$M_{ee}$ for these two solutions are tabulated in Table 1. Also,
it can be inferred from Fig. 1 that Dirac-type CP-violating phase
$\delta$ (Fig. 1(b)) and Majorana phase $\beta$ (Fig. 1(d)) remain
unconstrained for low as well as high $D_l$ solution. However,
Majorana phase $\alpha$ is constrained to the range $60^o-120^o$
for high $D_l$. In Fig. 2. we have plotted lightest neutrino mass
eigenvalue $m_1$ with $\theta_{13}$ for low (left) and high $D_l$
(right) region. This plot has been reported earlier \cite{16},
however, no constraints have been obtained on $D_l$ and
$\theta_{13}$. It is noted that the analysis done in \cite{16}
corresponds to low $D_l$ solution obtained in the present work. We
have obtained an upper bound on $m_1$ of the order of $10^{-2}$
for low as well as high $D_l$ solution which is in agreement with
that obtained in \cite{16}. Also, we find that $m_1$ is bounded
from below for low and high $D_l$ regions for $\theta_{13}<8^o$
and $<10^o$, respectively.

It should be noted that in performing numerical analysis we have
used value of $\theta_{13}$ obtained from global analysis of
neutrino data, given by Eq. (20). However, measurement of
$\theta_{13}$ is the main goal of the future experiments on
neutrino oscillations. The upcoming and proposed experiments in
this context include Double CHOOZ \cite{20,21}, Daya Bay
\cite{22,23} and RENO \cite{24}. Double CHOOZ is planned 
to explore sin$^22\theta_{13}$ down to $0.06$ in phase-I ($0.03$ in 
the later stages) and will be the first to report any observation on
$\theta_{13}$. The Daya Bay experiment has a higher sensitivity
and plans to observe $\sin^2 2\theta_{13}$ down to $0.01$. In this
light it will be useful to relax the constraint on $\theta_{13}$
[Eq. (20)] and examine the phenomenological implications of this
Ans\"{a}tz for small $\theta_{13}$. We have considered three
representative values of sin$^22\theta_{13}$ viz. $0.01$, $0.03$
and $0.06$ and examined the implications for the CP- violating
phases $\alpha$, $\beta$ and $\delta$ depicted in Fig. 3 and Fig.
4, respectively, for low and high $D_L$ solutions. It can be seen
from Fig. 3(a) and Fig. 3(b) that as $\sin^22\theta_{13}$ is
decreased, the Majorana phases $\alpha$, $\beta$ becomes more
constrained. Furthermore, Dirac-type CP-violating phase $\delta$
is constrained to narrower ranges as $\sin^2 2\theta_{13}$ is
lowered which is evident from Fig. 4(a). However, for high $D_l$
solution $\delta$ remain unconstrained. There is no solution for
low $D_l$ with sin$^22\theta_{13}=0.01$ in Fig. 3. and Fig. 4
because for low $D_l$ the third mixing angle $\theta_{13}>3^o$
which is, also, evident from Fig. 1(a).

\section{Conclusions}
In conclusion, we have considered a class of lepton mass matrices
with parallel texture structure in charged lepton and Majorana
neutrino mass matrices and obtained interesting constraints on the
parameters of lepton mass matrices. An interesting feature of this
analysis is the rather strong constraints obtained on
$\theta_{13}$ and the emergence of two classes of solutions
corresponding to 'low $D_l$' and 'high $D_l$' which is in contrast
to the analysis reported earlier \cite{16}. The implications of
these two classes have been examined for the unknown parameters of
the neutrino mass matrix like $\theta_{13}$, $M_{ee}$ and
CP-violating phases without the assumption of factorizable phases
in the neutrino mass matrix. Some of the predictions of this
analysis are testable in the forthcoming neutrino experiments.

\bigskip

\textit{\Large{Acknowledgments}}

 The research work of S. D. is supported
by the Board of Research in Nuclear Sciences (BRNS), Department of
Atomic Energy, Government of India \textit{vide} Grant No. 2004/
37/ 23/ BRNS/ 399. S. K., S. G. and S. V. acknowledges the
financial support provided by Council for Scientific and
Industrial Research (CSIR) and University Grants Commission (UGC),
Government of India, respectively.

\pagebreak

\begin{table}[b]
\begin{center}
 \begin{tabular}{|c|c|c|}
   \hline

   Parameter & Low $D_l$ & High $D_l$ \\
   \hline
   $\theta_{13}$ & $5.4^o-11.2^o$ & $3.4^o-11.2^o$ \\
   $D_l$ & ($-85-415$) MeV & ($1555-1665$) MeV \\
   $M_{ee}$ & $(1.0-1.5)\times10^{-3}$ eV & $(0.9-18.0)\times10^{-4}$ eV \\
      \hline
 \end{tabular}
 \end{center}
 \caption{The ranges of  $\theta_{13}$, $D_l$ and $M_{ee}$ at 95\% C.L..}
\end{table}

\begin{figure}
\begin{center}
\epsfig{file=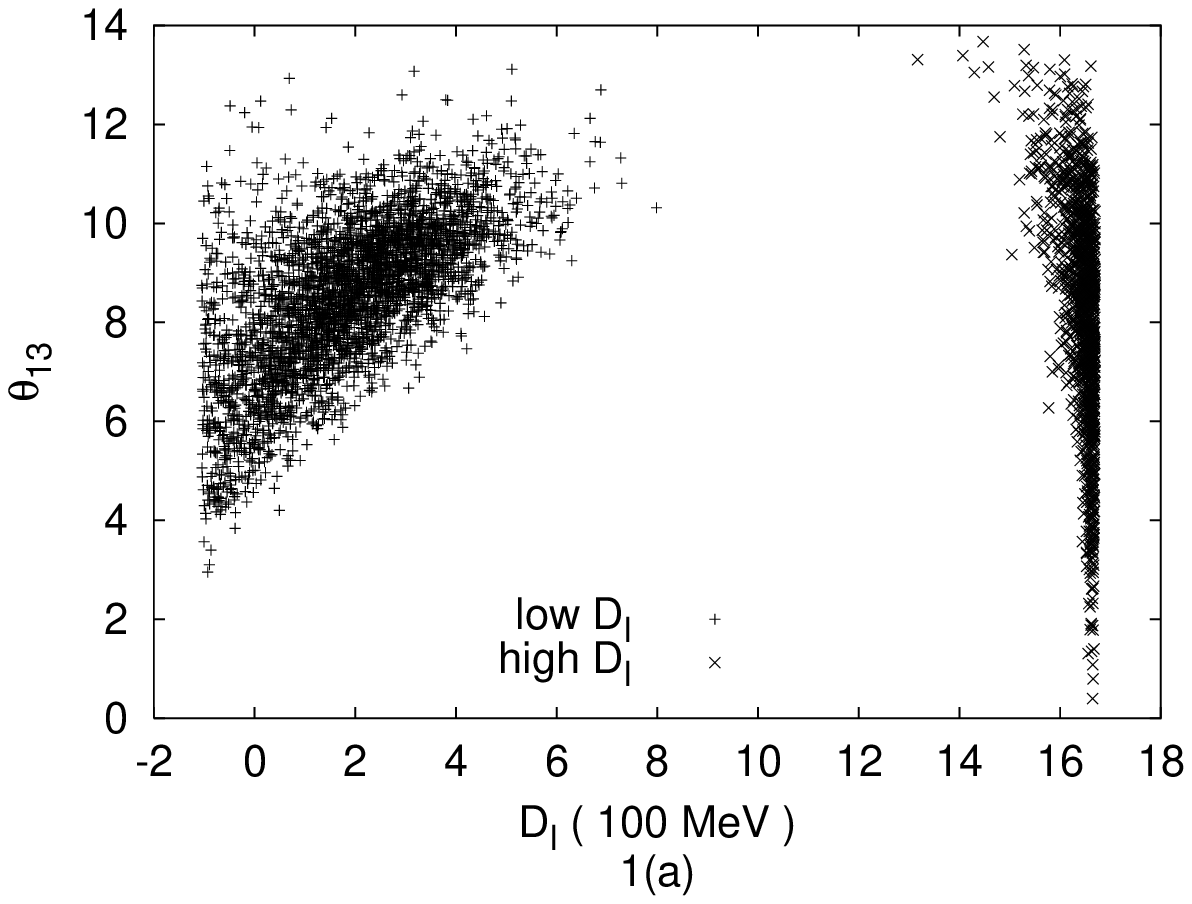, width=7.0cm, height=7.0cm}
\epsfig{file=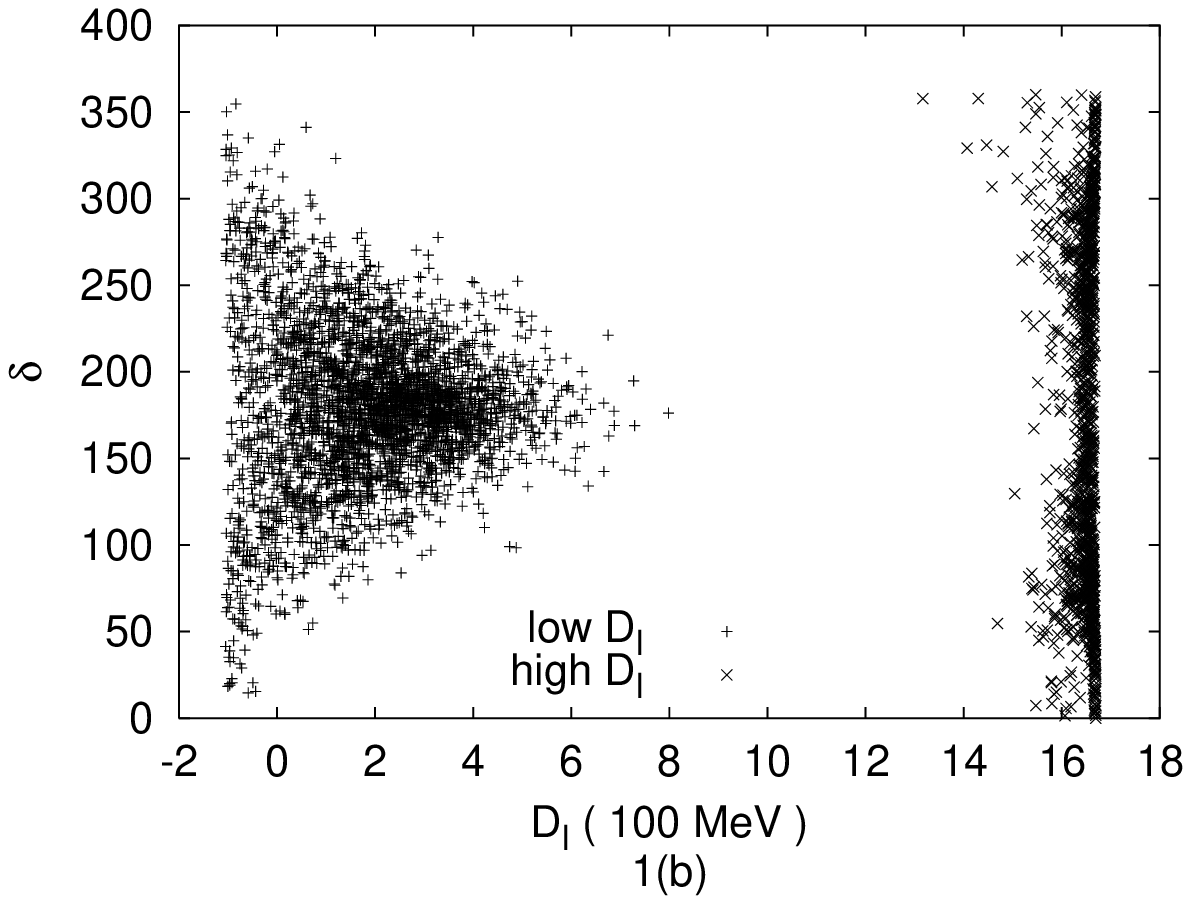, width=7.0cm, height=7.0cm}
\epsfig{file=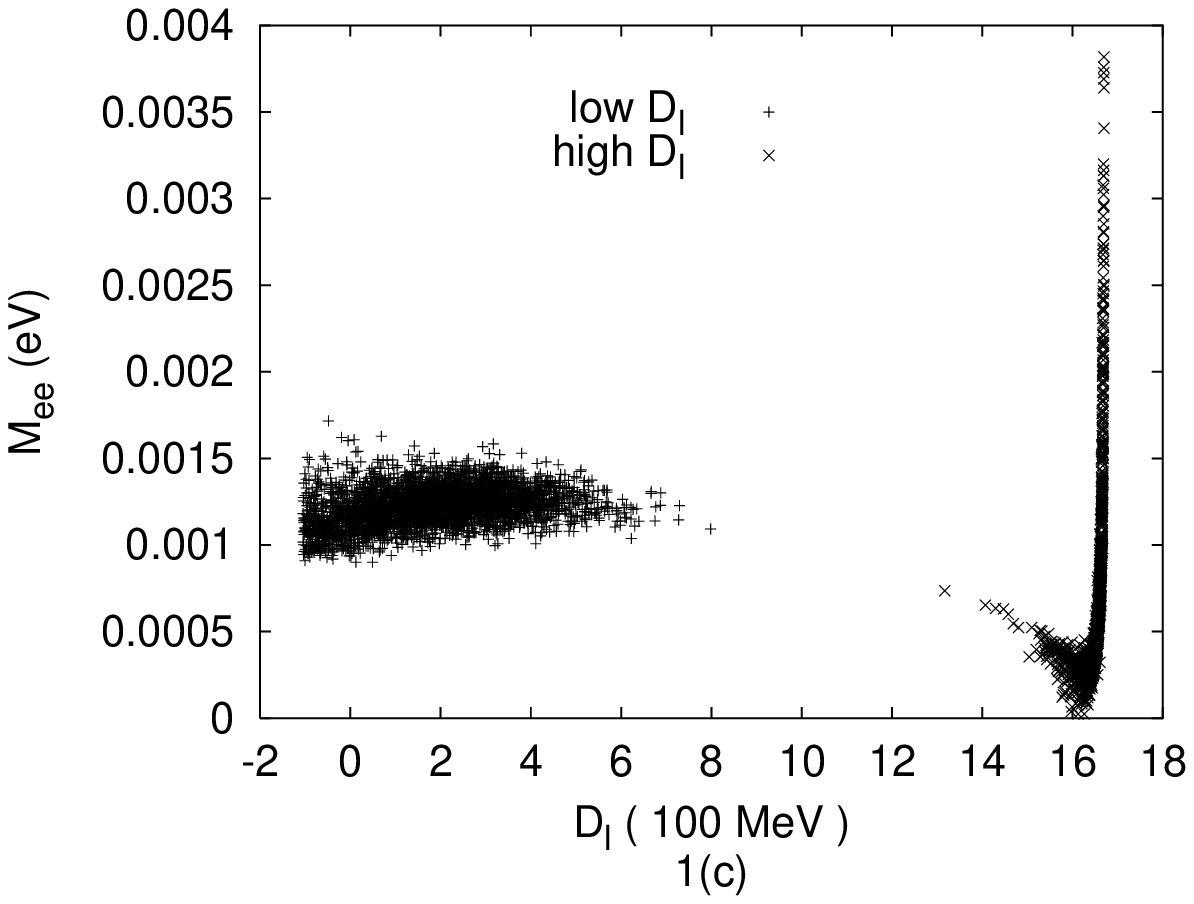, width=7.0cm, height=7.0cm}
\epsfig{file=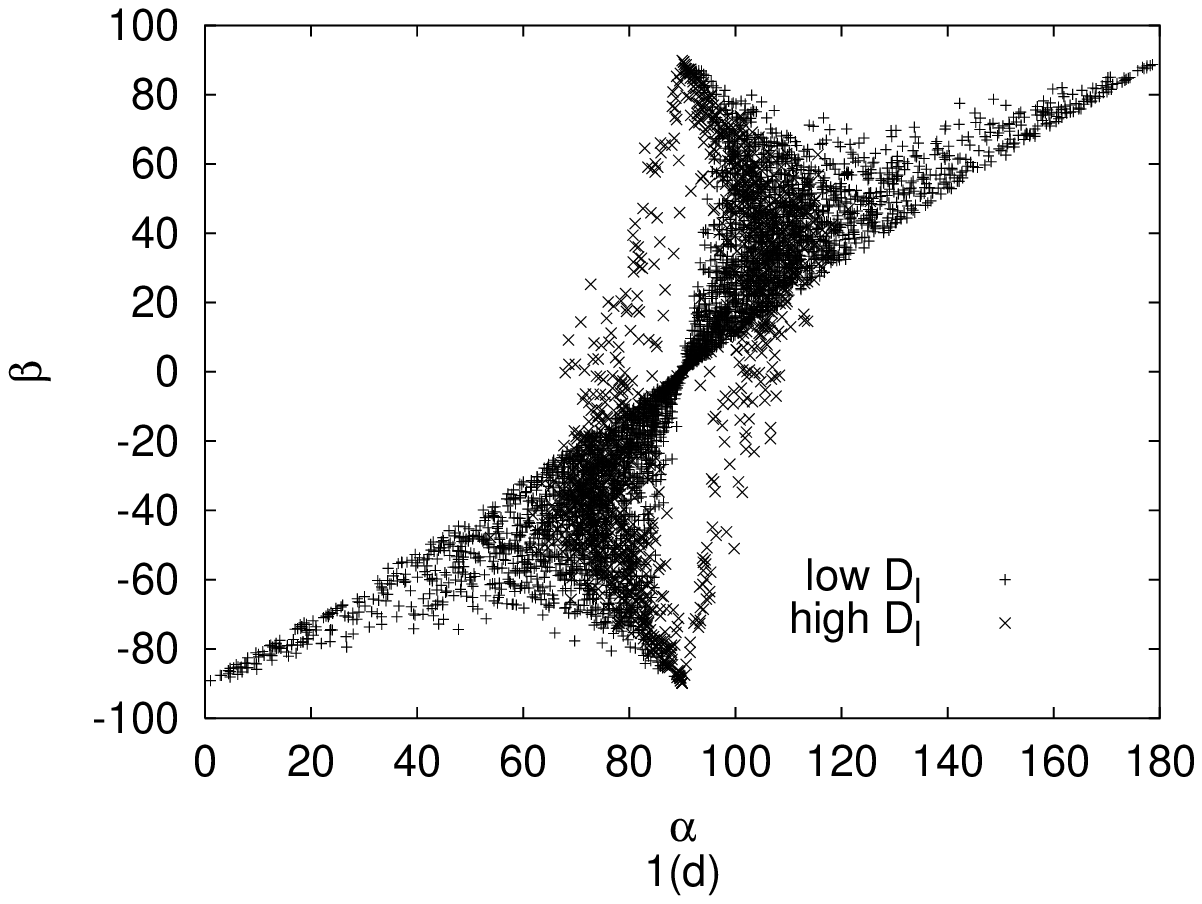, width=7.0cm, height=7.0cm}
\end{center}
\caption{The parameter space allowed by the current neutrino
oscillation data \cite{18} and $\theta_{13}$ from global analysis
\cite{19}.
 The scatter plots of $\theta_{13}$, $\delta$, $M_{ee}$ with $D_l$
and of $\beta$ with $\alpha$. Low $D_l$ (high $D_l$) solution is
represented by symbol 'plus' ('cross').}
\end{figure}

\begin{figure}
\begin{center}
\epsfig{file=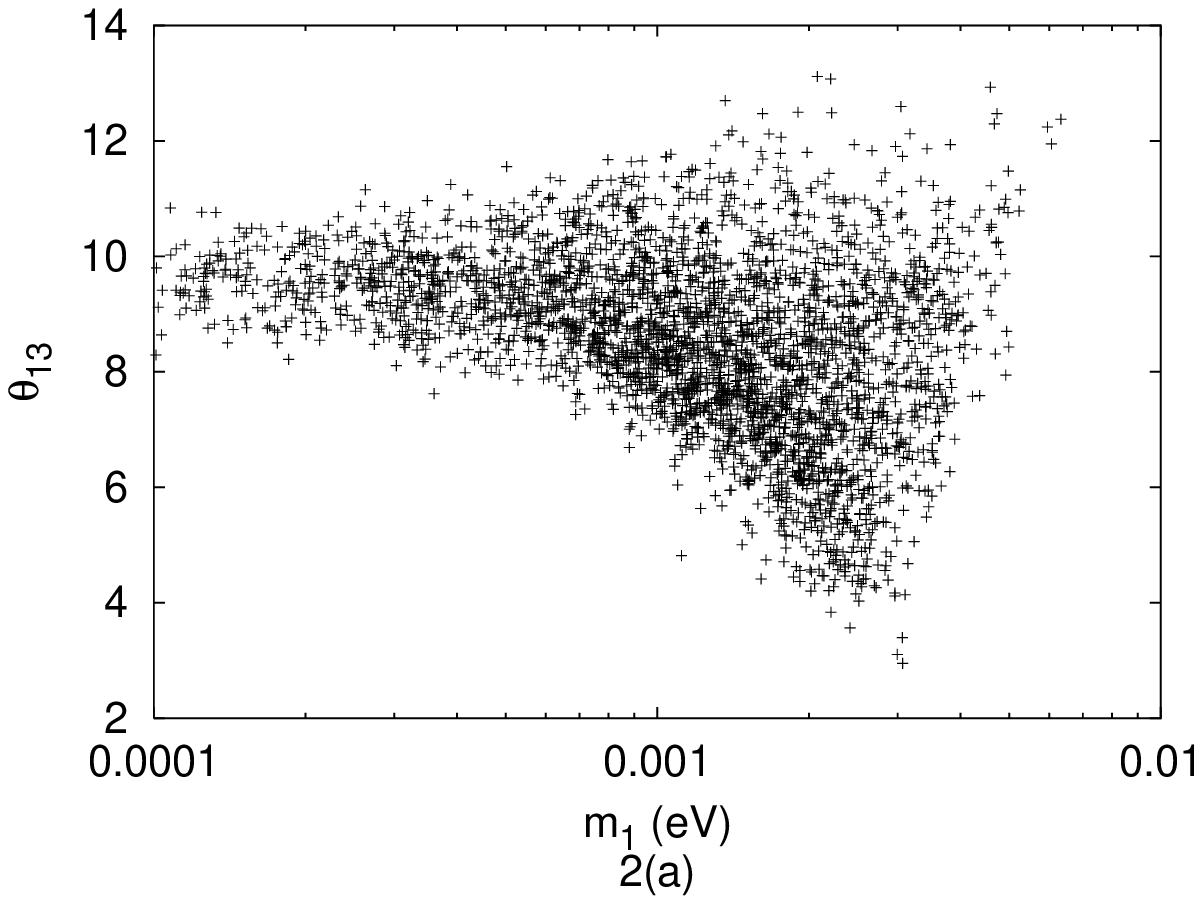, width=7.0cm, height=7.0cm}
\epsfig{file=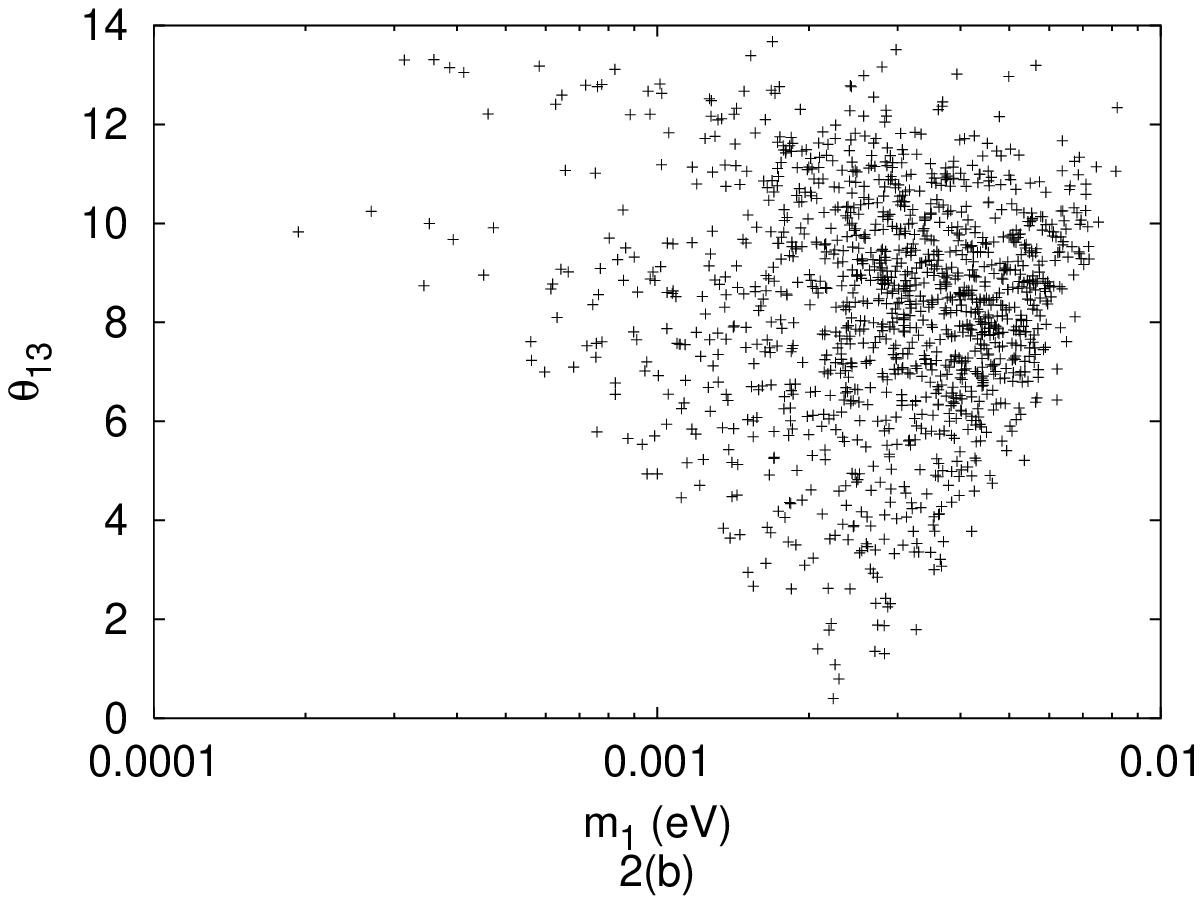, width=7.0cm, height=7.0cm}
\end{center}
\caption{($m_1-\theta_{13}$) correlation plots for low $D_l$
(left) and high $D_l$ (right).}
\end{figure}

\begin{figure}
\begin{center}
\epsfig{file=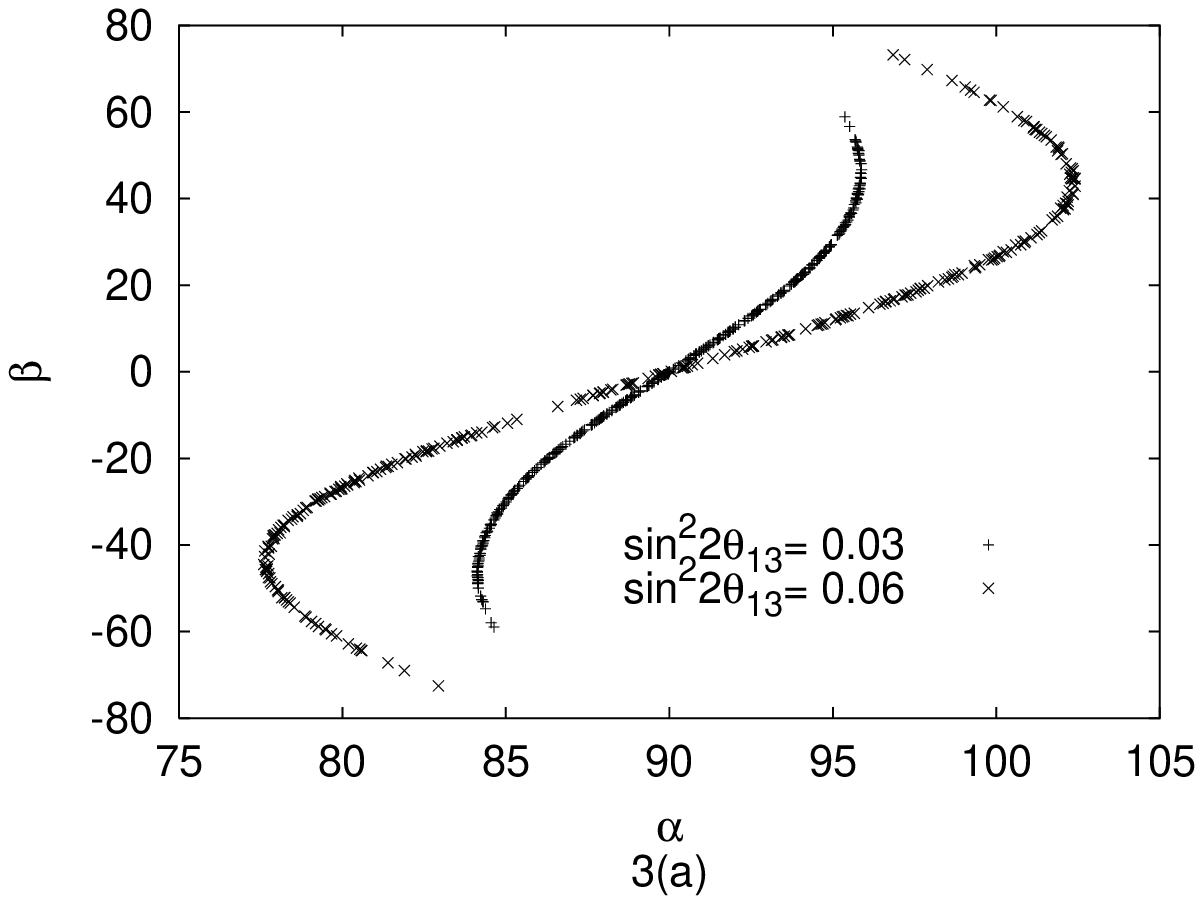, width=7.0cm, height=7.0cm}
\epsfig{file=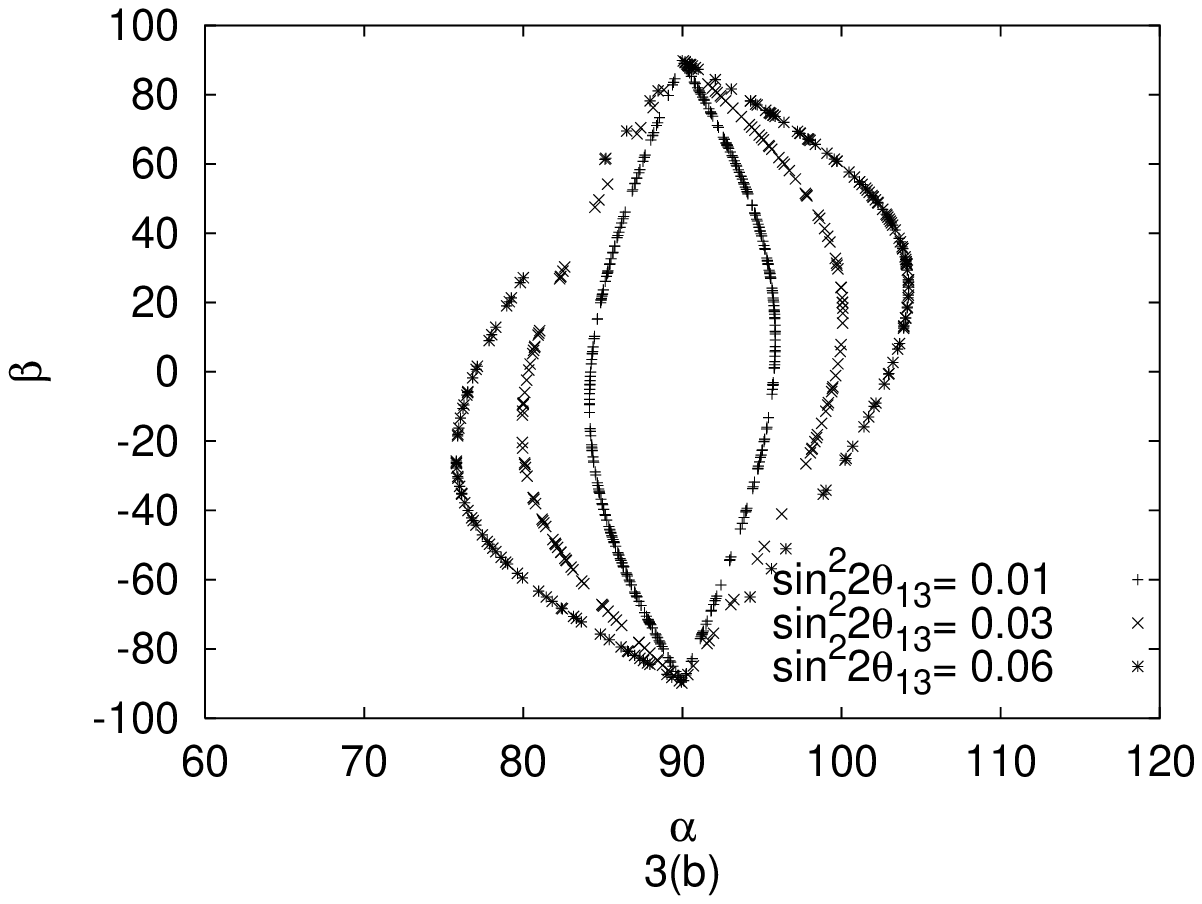, width=7.0cm, height=7.0cm}
\end{center}
\caption{Correlation plot of the allowed points in
($\alpha-\beta$) plane for low $D_l$ (left) and high $D_l$ region
(right).}
\end{figure}
\pagebreak{}

\begin{figure}
\begin{center}
\epsfig{file=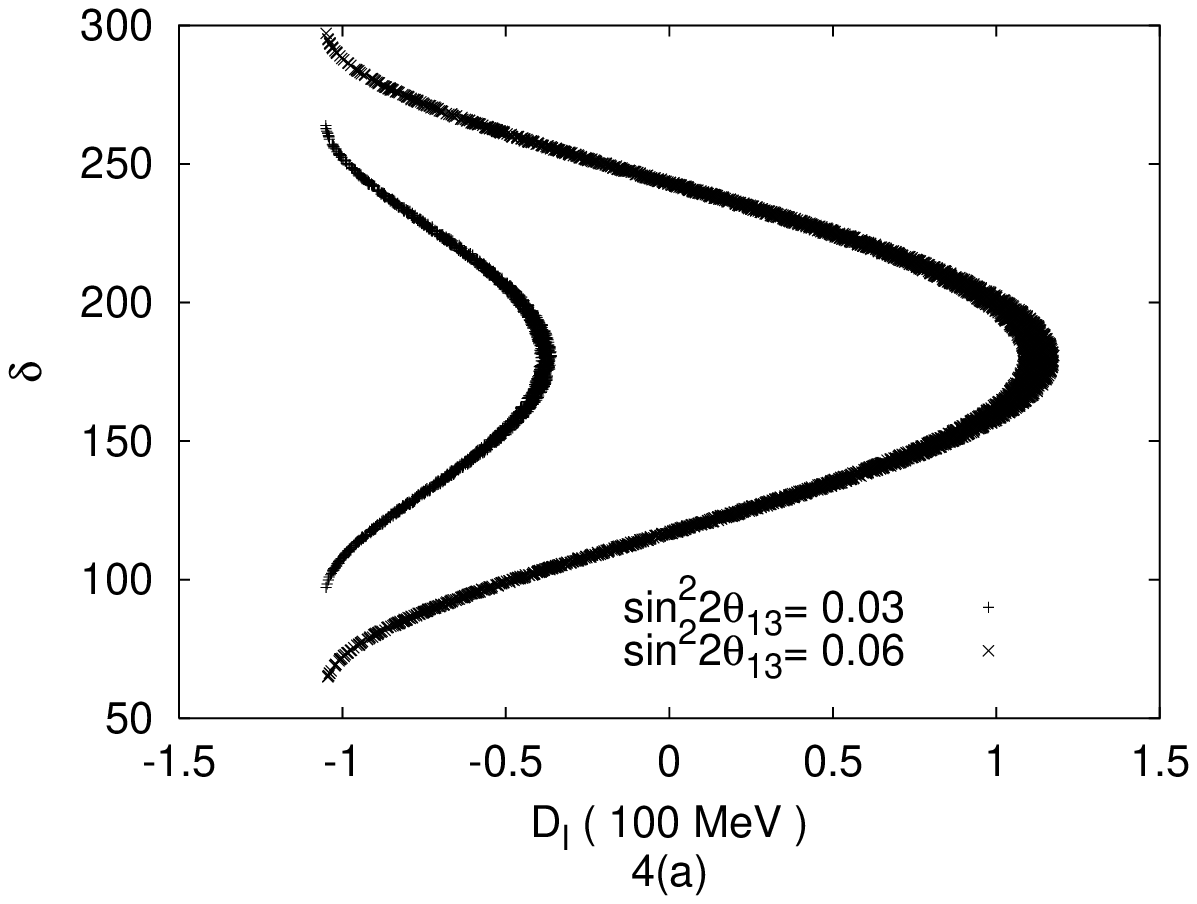, width=7.0cm, height=7.0cm}
\epsfig{file=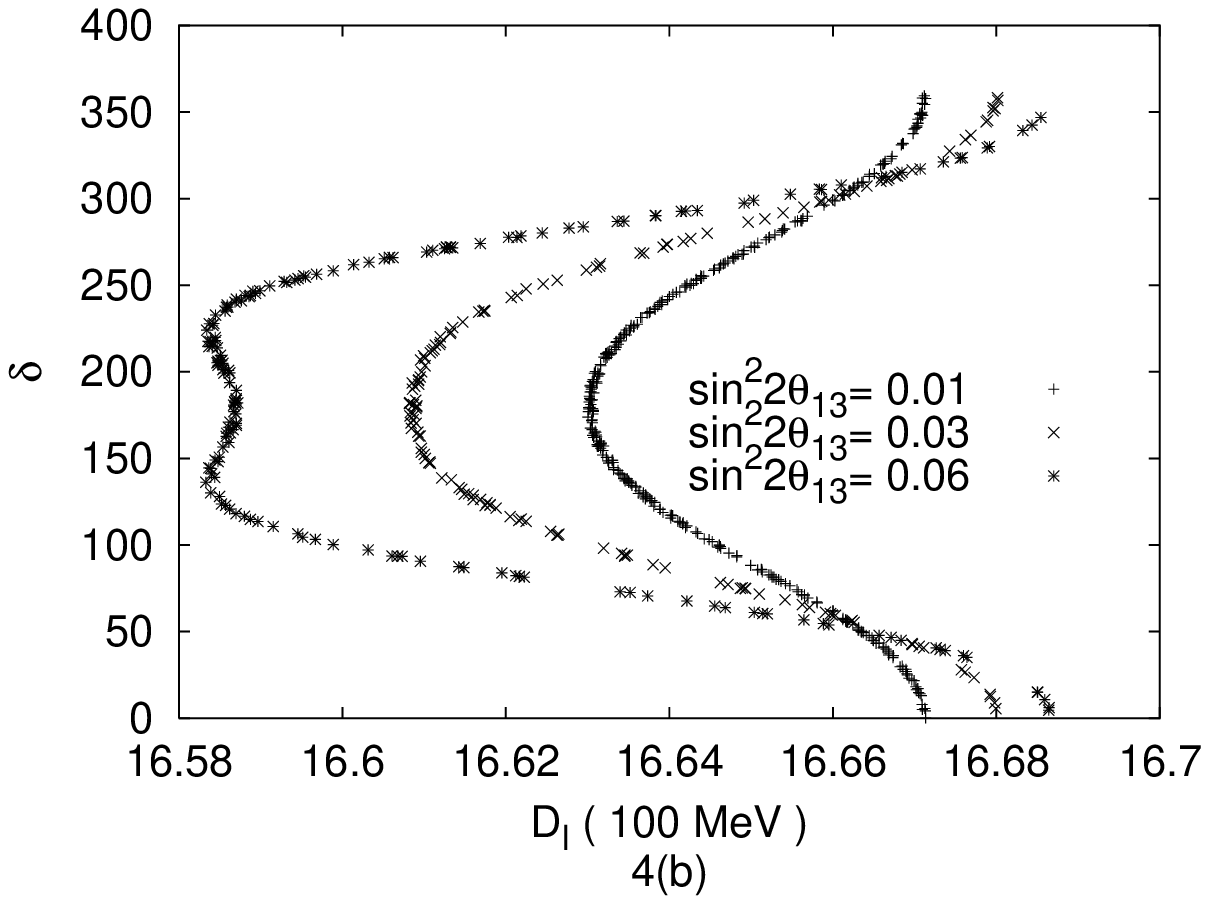, width=7.0cm, height=7.0cm}
\end{center}
\caption{($D_l-\delta$) correlation plot for
sin$^22\theta_{13}=0.01$, sin$^22\theta_{13}=0.03$ and
sin$^22\theta_{13}=0.06$ in low (left) and high (right) $D_l$
region.}
\end{figure}


\begin{thebibliography}{99}

\bibitem{1} Paul H. Frampton, Sheldon L. Glashow and Danny
Marfatia, \textit{Phys. Lett.} \textbf{B 536}, 79 (2002).

\bibitem{2} Bipin R. Desai, D. P. Roy and Alexander R. Vaucher, \textit{Mod. Phys. Lett} \textbf{A 18}, 1355 (2003).

\bibitem{3}  Zhi-zhong Xing, \textit{Phys. Lett.} \textbf{B 530},
159 (2002).

\bibitem{4} Wanlei Guo and Zhi-zhong Xing, \textit{Phys. Rev.} \textbf{D 67}, 053002 (2003).

\bibitem{5} Alexander Merle and Werner Rodejohann, \textit{Phys. Rev.} \textbf{D 73}, 073012 (2006). 

\bibitem{6} S. Dev and Sanjeev Kumar, hep-ph/0607048.

\bibitem{7} S. Dev, Sanjeev Kumar, Surender Verma and Shivani 
Gupta, \textit{Nucl. Phys.} \textbf{B 784}, 103 (2007).

\bibitem{8} S. Dev, Sanjeev Kumar, Surender Verma and Shivani 
Gupta, \textit{Phys. Rev.} \textbf{D 76}, 013002 (2007).

\bibitem{9}  G. C. Branco, R. Gonzalez Felipe, F. R. Joaquim and T. 
Yanagida, \textit{Phys. Lett.} \textbf{B 562} 265 (2003).

\bibitem{10} Bhag C. Chauhan, Joao Pulido and Marco Picariello, 
\textit{Phys. Rev.} \textbf{D 73}, 053003 (2006).

\bibitem{11} Xiao-Gang He and A. Zee, \textit{Phys. Rev.} \textbf{D 68}, 037302 (2003), hep-ph/0302201 v2.

\bibitem{12} Monika Randhawa, V. Bhatnagar, P. S. Gill, M. Gupta,
\textit{Phys. Rev.} \textbf{D 60}, 051301 (2005).

\bibitem{13} Koichi Matsuda, Hiroyuki Nishiura, 
 \textit{Phys. Rev.} \textbf{D 74}, (2006) 033014.

\bibitem{14} Gulsheen Ahuja, Sanjeev Kumar, Monika Randhawa,
Manmohan Gupta, S. Dev, \textit{Phys. Rev.} \textbf{D 76}, 013006
(2007).

\bibitem{15} Z. z. Xing and H. Zhang, \textit{Phys. Lett.} \textbf{B
569}, 30 (2003).

\bibitem{16} G. C. Branco, D. Emmanuel-Costa, R. Gonz\'{a}lez 
Felipe and H.Ser\^{o}dio, \textit{JHEP} \textbf{0801}, 043 (2008), 
hep-ph/0711.1613v1.

\bibitem{17} C. Amsler \textit{et al}., \textit{Phys. Lett.} \textbf{B
667}, 1 (2008).

\bibitem{18} M. C. Gonzalez-Garcia, Michele Maltoni, \textit{Phys. Rept.} 460 (2008) 1-129, hep-ph/0704.1800v2.

\bibitem{19} G. L. Fogli, E. Lisi, A. Marrone, A. Palazzo,
A. M. Rotunno, arXiv:hep-ph/0806.2649.

\bibitem{20} http://doublechooz.in2p3.fr/.

\bibitem{21} F. Ardellier \textit{et al.} (Double Chooz Collaboration), hep-ex/0606025.

\bibitem{22} http://dayabay.ihep.ac.in/.

\bibitem{23} Daya Bay Collaboration, hep-ex/0701029.

\bibitem{24} http://neutrino.snu.ac.kr/RENO/.

\end{thebibliography}
\end{document}